\documentclass[%
 reprint,
 amsmath,amssymb,
 aps,
]{revtex4-2}

\usepackage{graphics,tikz}
\usetikzlibrary{shapes.misc,arrows,decorations.markings,patterns,calc,shapes.geometric,intersections}
\usepackage{amsmath, amssymb, graphics}
\usepackage{graphicx}
\usepackage{dcolumn}
\usepackage{bm}
\usepackage[
    colorlinks=true,
    urlcolor=blue,
    linkcolor=blue,
    citecolor=blue
]{hyperref}

\usepackage{color}
\usepackage{braket}
\usepackage{comment}

\newcommand{\del}{\partial}

\def\Dm{{\mathcal{D}}}

\def\Nm{{\mathcal{N}}}
\def\Om{{\mathcal{O}}}

\def\Sm{{\mathcal{S}}}

\def\Xm{{\mathcal{X}}}

\definecolor{UnitoRed}{HTML}{EA0029}
\definecolor{LightGrey}{HTML}{f4f4f4}

\newcommand{\lsp}{\hspace{0.5pt}}

\newcommand\mm[1]{{\color{red} \textbf{mm:  #1}}}

\begin{document}

\preprint{APS/123-QED}

\title{Conformal defects and Goldstone bosons in Anti-de Sitter space}

\author{Lorenzo Bianchi}
\email{lorenzo.bianchi@unito.it}
 \affiliation{Dipartimento di Fisica, Universit\`a di Torino and INFN - Sezione di Torino\\ Via P. Giuria 1, 10125 Torino, Italy}
 
\author{Elia de Sabbata}%
 \email{eliadesabbata@gmail.com}
\affiliation{%
 Department of Mathematics, King's College London, \\
Strand, London, WC2R 2LS, United Kingdom 
}%

\author{Marco Meineri}
 \email{marco.meineri@unito.com}
\affiliation{Dipartimento di Fisica, Universit\`a di Torino and INFN - Sezione di Torino\\ Via P. Giuria 1, 10125 Torino, Italy
}%

\begin{abstract}

We study local quantum field theories in Anti-de Sitter (AdS) space, with boundary conditions that break some of the bulk isometries. Specifically, we focus on conformal defects and we prove that their spectrum supports a displacement operator of protected dimension, despite the non-local nature of the conformal theory living at the boundary of AdS. If the defect breaks a global symmetry, a tilt operator is also present. The existence of a displacement was conjectured in \cite{Gabai:2025hwf} for Wilson loops in Yang-Mills theories in AdS. Our proof is valid in general and applies, in particular, to defects in long-range models, as we discuss in various examples. In the bulk, the modes sourced by the protected operators have Compton wavelength of order of the AdS radius: they constitute the AdS analogue of the Goldstone bosons for the spontaneous breaking of the corresponding symmetries.

\end{abstract}

\maketitle

\section{\label{sec:level1}Introduction}

The study of quantum field theory (QFT) in hyperbolic space has seen a remarkable development, since the dream of using curvature as an infrared (IR) regulator to study strongly coupled physics \cite{CALLAN1990366} was married to the conformal bootstrap \cite{Paulos:2016fap} and large-$N$ technologies \cite{Carmi:2018qzm}. When the Anti-de Sitter (AdS) \footnote{While for simplicity we work in Euclidean signature, nothing crucially depends on it, so we will not distinguish between hyperbolic space and AdS.}  radius $R$ is larger than any other scale, physics in AdS smoothly connects to flat space dynamics, where some of the most interesting questions have to do with low energy properties of asymptotically free theories: spontaneous symmetry breaking (SSB) or symmetry restoration \cite{Coleman1973}, confinement and chiral symmetry breaking \cite{PhysRevD.10.2445}. Many of these phases are distinguished by the presence (or absence) of massless excitations: the Goldstone bosons \cite{Goldstone1961} associated to the breaking of global or spacetime symmetries. In AdS, on the other hand, the radius regulates the IR and gaps the dynamics. Furthermore, the boundary conditions do not decouple. Vacua where the order parameter has a vev correspond to boundary conditions which break the symmetry explicitly, and one may wonder what feature of SSB persists as the radius is varied.

In the case of continuous global symmetries, this question was clarified thanks to the connection between QFT in AdS and defect conformal field theory (dCFT) \cite{McAvity:1995zd,Billo:2016cpy}. When a conformal defect breaks a continuous symmetry of the CFT, a protected \emph{tilt} operator exists on the defect \cite{AJBray_1977}. Correspondingly, the boundary of AdS$_{d+1}$ supports a tilt operator with dimension $\Delta=d$, which, in the large $R$ limit, sources a single Goldstone boson \cite{Copetti:2023sya}. At finite radius, the Goldstone bosons are strongly interacting, and the existence of the tilt and the identities obeyed by its correlation functions \cite{Bianchi:2018zpb,Padayasi:2021sik,Cavaglia:2022qpg,Drukker:2022pxk,Gabai:2025zcs,Girault:2025kzt,Drukker:2025dfm} provide the non-perturbative hallmark of SSB.

Confining theories arguably represent the most attractive challenge for QFT in AdS. In flat space, Wilson loops exhibit area law and source a string whose transverse fluctuations are Goldstone bosons for the breaking of isometries \cite{Luscher:1980ac}. In AdS, the area law fails to diagnose confinement \cite{CALLAN1990366}, but the flux tube sourced by a Wilson loop anchored at the boundary exists at all radii due to the gravitational potential, opening the enticing possibility that perturbative computations might capture flat space spectra and Wilson coefficients \cite{Gabai:2025hwf,Gabai:2026myo}. In \cite{Gabai:2025hwf}, it was conjectured that a protected \emph{displacement} operator exists on the Wilson line, based on two complementary perspectives: at small $R$, the explicit definition of the line as the holonomy of the gauge field; at large $R$, the assumption that a two-dimensional description of the flux tube becomes valid. The displacement operator has $\Delta=2$ \cite{TWBurkhardt_1987,Billo:2013jda}, and sources a Goldstone boson which propagates on the worldsheet in the large radius limit. At intermediate radii, the flux tube spreads over a size $R$, and a worldsheet description is not available at the scale of the mass of the boson. Conversely, the standard argument for the existence of the displacement \cite{Billo:2016cpy} does not apply because the conformal theory at the boundary of AdS lacks a stress tensor. 

Nevertheless, the main purpose of this paper is to prove that the displacement exists, at any value of $R$. The proof is similar in spirit to the flat space one \cite{PhysRev.127.965}, based on the spectral density of the two-point function of the order parameter and the current. It is valid for any conformal defect on the boundary of AdS, and in particular it does not assume a Lagrangian, nor any particular description of the field configuration in AdS. With minor modifications, our argument also applies to the breaking of continuous global symmetries---which imply the existence of a tilt operator on the defect---and of higher spin symmetries.

The rest of the paper is organised as follows. Our main result is derived in section \ref{sec:generalargument}, followed by some examples in section \ref{sec:examples} and conclusions in section \ref{sec:conclusions}. The End Matter, sections \ref{section:DOEstress}-\ref{section:q=1}, are dedicated to useful technical details. In the Supplemental Material, we describe the consequences of SSB of global symmetries (\ref{tilt}), more details on the examples mentioned in the main text (\ref{sec:examplesSM}), and, for completeness, some consequences of the conservation of the stress tensor which go beyond what is needed in the main text (\ref{section:conservationequations}).

Several consequences of the main statement are explored in the conclusions, including its validity for defects in generic non-local CFTs and its implications on the properties of defect conformal manifolds. The role of the displacement as an order parameter for confinement is also further discussed.  \\

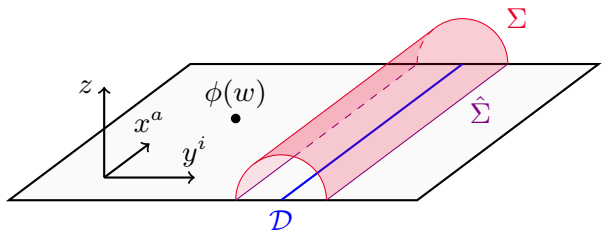
\begin{figure}
    \centering
    \begin{tikzpicture}[scale=1.2, transform shape]


   
    \coordinate (A) at (-0.5,0);
    \coordinate (B) at (4,0);
    \coordinate (C) at (6,1.5);
    \coordinate (D) at (1.5,1.5);

    \fill[LightGrey, opacity=0.4] (A) -- (B) -- (C) -- (D) -- cycle;
    \draw[black,thick] (A) -- (B) -- (C) -- (D) -- cycle;

    \coordinate (P) at (2.5,0);
    \coordinate (Q) at (4.5,1.5);
    \draw[blue, thick] (P) -- (Q);

    \coordinate (O) at (0.55,0.25);
    \draw[->, thick] (O) -- ++(1,0) node[above] {$y^i$};
    \draw[->, thick] (O) -- ++(0.5,0.375) node[above] {$x^a$};
    \draw[->, thick] (O) -- ++(0,1) node[left] {$z$};

    \path (3,0) arc[start angle=0,end angle=127,radius=0.5] coordinate (E);
    \path (5,1.5) arc[start angle=0,end angle=127,radius=0.5] coordinate (F);

    \path[name path=line] (2,0) -- (4,1.5);
    \path[name path=semicirc] (3,0) arc (0:180:0.5);
    \path[name intersections={of=line and semicirc, by={H,I}}];

    \draw[dashed,violet] (H) -- (4,1.5);
    \draw[UnitoRed] (5,1.5) arc (0:127:0.5);
    \draw[dashed,UnitoRed] (4,1.5) arc (180:127:0.5);
    \draw[UnitoRed] (3,0) arc (0:180:0.5);
    \draw[violet] (3,0) -- (5,1.5);
   
    \pgfmathsetmacro{\m}{0.75}
    \pgfmathsetmacro{\q}{-0.97+(-30+5*sqrt(26))/16}
    \draw[UnitoRed,domain=2.2:4.2] plot (\x, {\m*\x + \q});
    \node[UnitoRed] at (5.1,2) {$\Sigma$};
    \node[blue,below] at (2.5,0) {$\mathcal{D}$};
    \draw[violet] (2,0) -- (H);  
    \node[violet] at (4.7,1) {$\hat\Sigma$};

    \fill[UnitoRed, opacity=0.1] (2,0) arc[start angle=180,end angle=127,radius=0.5] -- (E) -- (F) arc[start angle=127,end angle=180,radius=0.5] -- (4,1.5) -- cycle;;

    \fill[UnitoRed, opacity=0.2]
    (E) arc[start angle=127,end angle=0,radius=0.5] -- (5,1.5)
    arc[start angle=0,end angle=127,radius=0.5] -- cycle;

    \fill (2,0.9) circle (1.5pt) node[above] {$\phi(w)$};

    \end{tikzpicture}    
    \caption{The charge $Q_{\xi}$ is integrated over $\Sigma$ surrounding the defect. The operator $\phi$ lies on the AdS boundary: for brevity, we use the notation $\phi(w)$ with the understanding that in this case $w^\mu=(x^a,y^i,0)$.}
    \label{fig:tunnel}
\end{figure}

\paragraph*{\textbf{Note added.}}
While finalizing this manuscript, the authors became aware of closely related work by Jiaxin Qiao, who independently obtained the same results with similar arguments in \cite{Qiao:2026}. Following mutual communication, we coordinated the submission so that the two papers appear on the same date.

\section{Conformal charges in AdS and the Goldstone theorem}
\label{sec:generalargument}

Consider a \emph{local} QFT---by which we mean a theory endowed with a stress tensor---on a fixed $(d+1)$-dimensional AdS background. The $d$-dimensional CFT at the AdS boundary is non-local.
A conformal defect, located on a flat $p$-dimensional surface $\Dm$ on the boundary $\partial \text{AdS}$, breaks the AdS isometries to the defect symmetry group $SO(p+1,1)\times SO(q)$, with $p+q=d$. In a \emph{local} $d$-dimensional CFT, a standard argument \cite{Billo:2016cpy} shows the existence of a \emph{displacement} operator $\hat D_i$ --- $i=1,\dots q$ --- with dimension $\hat\Delta=p+1$, transforming as a vector under the transverse rotation group $SO(q)$.

To show that the same happens on $\partial \text{AdS}$, we will use the fact that the conformal charges have a local expression in terms of the bulk stress tensor. The action of the charge $Q_\xi$, associated to the conformal Killing vector $\xi$, on boundary operators enclosed by the surface $\hat{\Sigma}$, is
\begin{equation}
Q_\xi=\int_{\Sigma} d\Sigma_{\mu}\xi_\nu T^{\mu \nu}~, 
\label{Qxi}
\end{equation}
where $\Sigma$ is an open codimension-one surface in the AdS interior, whose boundary $\hat{\Sigma}$ lies in $\partial AdS$---see figure \ref{fig:tunnel}. As in the flat space story \cite{PhysRev.127.965}, it is useful to consider the spectral decomposition of the two-point function $\braket{\phi T_{\mu\nu}}$ of the current with an order parameter, \emph{i.e.}\ an operator $\phi$ whose expectation value in the presence of the defect $\braket{\phi}$ does not vanish (hereafter, the presence of the defect will always be understood in correlators). By choosing the quantization scheme appropriately, the spectrum in AdS is labeled by scaling operators on the defect. The action of the charges \eqref{Qxi} on local operators constrains the spectral density, and the constraints can only be satisfied if the defect spectrum includes a displacement multiplet. The displacement obeys the usual integrated constraints \cite{Billo:2016cpy}, which are encapsulated in the following Ward identity
\begin{equation}\label{dispDiffWard}
   \sqrt{g}\, \nabla^\mu T_{\mu i}(w) =\sqrt{\mathcal{C}_{\hat{D}}}\, \delta^{q}(y)\delta(z)\, \hat D_i(x)\,,
\end{equation}
where $\mathcal{C}_{\hat{D}}$ is a physical constant, provided that the two-point function of $\hat{D}_i$ is unit-normalized, as we will assume of all operators except the stress tensor, unless otherwise stated.  For clarity, we chose Poincaré coordinates 
$w^\mu = (x^a, y^i,z)$ 
and split them among parallel to $\mathcal{D}$, perpendicular to it on $\partial$AdS, and orthogonal to $\partial$AdS.

The proof relies on unitarity, and crucially, on the topological dependence of the charges on the surface $\Sigma$, including its boundary $\hat{\Sigma}$. In the bulk, the condition is violated by defects which extend in the AdS interior, like non-dynamical branes \cite{Karch:2000gx}, thus breaking the isometries explicitly. On $\partial$AdS, the independence of the charges from the local shape of $\hat{\Sigma}$ is violated, for instance, by turning on background field configurations which break $SO(d+1,1)$ to $SO(p+1,1)\times SO(q)$ away from the defect \cite{Herzog:2019bom}. As we describe in section \ref{sec:examples}, in both cases the displacement is absent.

In the rest of the section, we describe the proof in detail.

\subsection{The stress tensor and the displacement operator}
\label{sec:disp}

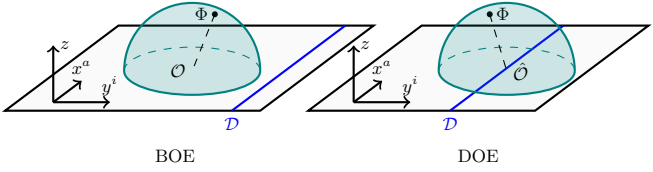
\begin{figure}[htbp]
\centering
\begin{tikzpicture}[scale=0.75, transform shape]

```

\coordinate (A) at (-0.5,0);
\coordinate (B) at (4,0);
\coordinate (C) at (6,1.5);
\coordinate (D) at (1.5,1.5);

\fill[LightGrey, opacity=0.4] (A) -- (B) -- (C) -- (D) -- cycle;
\draw[black,thick] (A) -- (B) -- (C) -- (D) -- cycle;

\coordinate (P) at (3.5,0);
\coordinate (Q) at (5.5,1.5);
\draw[blue, thick] (P) -- (Q);

\coordinate (O) at (0.35,0.15);
\draw[->, thick] (O) -- ++(1,0) node[above] {$y^i$};
\draw[->, thick] (O) -- ++(0.5,0.375) node[above] {$x^a$};
\draw[->, thick] (O) -- ++(0,1) node[right] {$z$};

\node[blue,below] at (3.5,0) {$\mathcal{D}$};

\fill (3.2,1.7) circle (1.5pt) node[left] {$\Phi$};

\begin{scope}
    \coordinate (S) at (2.8,0.7);
    \def\r{1.2}

    \fill[teal, opacity=0.18]
        (S) ++(-\r,0)
        arc[start angle=180, end angle=0, x radius=\r, y radius=\r]
        arc[start angle=0, end angle=-180, x radius=\r, y radius=0.35*\r]
        -- cycle;

    \draw[teal, thick]
        (S) ++(-\r,0)
        arc[start angle=180, end angle=0, x radius=\r, y radius=\r];

    \draw[teal, thick]
        (S) ++(\r,0)
        arc[start angle=0, end angle=-180, x radius=\r, y radius=0.35*\r];

    \draw[teal, dashed]
        (S) ++(-\r,0)
        arc[start angle=180, end angle=0, x radius=\r, y radius=0.35*\r];

    \draw[dashed] (3.2,1.7)--(S);

    \node[left] at (S) {$\mathcal{O}$};
\end{scope}
\node at (2.5,-0.8) {BOE};

\end{tikzpicture}\hspace{-1cm}
\begin{tikzpicture}[scale=0.75, transform shape]

```

\coordinate (A) at (0,0);
\coordinate (B) at (4,0);
\coordinate (C) at (6,1.5);
\coordinate (D) at (2,1.5);

\fill[LightGrey, opacity=0.4] (A) -- (B) -- (C) -- (D) -- cycle;
\draw[black,thick] (A) -- (B) -- (C) -- (D) -- cycle;

\coordinate (P) at (2.5,0);
\coordinate (Q) at (4.5,1.5);
\draw[blue, thick] (P) -- (Q);

\coordinate (O) at (0.8,0.15);
\draw[->, thick] (O) -- ++(1,0) node[above] {$y^i$};
\draw[->, thick] (O) -- ++(0.5,0.375) node[above] {$x^a$};
\draw[->, thick] (O) -- ++(0,1) node[right] {$z$};

\node[blue,below] at (2.5,0) {$\mathcal{D}$};

\fill (3.2,1.7) circle (1.5pt) node[right] {$\Phi$};

\begin{scope}
    \coordinate (S) at (3.5,0.7);
    \def\r{1.2}

    \fill[teal, opacity=0.18]
        (S) ++(-\r,0)
        arc[start angle=180, end angle=0, x radius=\r, y radius=\r]
        arc[start angle=0, end angle=-180, x radius=\r, y radius=0.35*\r]
        -- cycle;

    \draw[teal,thick]
        (S) ++(-\r,0)
        arc[start angle=180, end angle=0, x radius=\r, y radius=\r];

    \draw[teal,thick]
        (S) ++(\r,0)
        arc[start angle=0, end angle=-180, x radius=\r, y radius=0.35*\r];

    \draw[teal,dashed]
        (S) ++(-\r,0)
        arc[start angle=180, end angle=0, x radius=\r, y radius=0.35*\r];

    \draw[dashed] (3.2,1.7)--(S);
    \node[right] at (S) {$\hat{\mathcal{O}}$};
    
\end{scope}
\node at (3,-0.8) {DOE};

\end{tikzpicture}

\caption{The two quantization schemes we consider.}
\label{fig:quantizations}

\end{figure}

It will be useful to distinguish two ways of slicing AdS. In Poincaré coordinates, we can pick a constant slice to be a half-sphere centered on a point away from the defect, or on a point belonging to it (see Figure \ref{fig:quantizations}). Quantizing on the first kind of slices, one concludes that a product of boundary operators can be replaced with a sum over scaling boundary operators (Operator Product Expansion, OPE), and a bulk operator can be replaced with a sum of boundary operators (Boundary operator Expansion, BOE). The second kind of slicing proves the convergence of the Defect Operator Expansion (DOE), which replaces a bulk or boundary operator with a sum of excitations of the defect.

Now, let $\phi$ be a scaling operator, other than the identity, on the boundary of AdS, such that $\braket{\phi}\neq 0$. At least one such operator must exist, for otherwise the defect would be disconnected from any correlators of AdS or $\partial$AdS operators, as one can conclude by repeated application of the BOE and of the OPE. Then, consider $\braket{Q_\xi \phi}$, where the charge \eqref{Qxi} is defined over the surface $\Sigma$ depicted in figure \ref{fig:tunnel}. Its boundary can be chosen as $\hat{\Sigma}=S^{q-1}\times \mathbb{R}^p$. By deforming the integration region outward, one concludes $\braket{Q_\xi \phi}=\del_\xi\braket{ \phi}\neq 0$, for an AdS Killing vector belonging to $SO(d+1,1)/SO(q)\times SO(p+1,1).$

One can also evaluate $\braket{Q_\xi \phi}$ using the DOE of the stress tensor, \emph{i.e.}\ compute the spectral decomposition of $\braket{\phi T_{\mu\nu}}$. It is important that the DOE commutes with the integral in \eqref{Qxi}, a fact that follows from convergence of the DOE and BOE and that we prove in section \ref{section:DOEforQ} in the End Matter. Defect operators are labeled by representations of $SO(p)\times SO(q)$ and we focus on the ones that can contribute to the charge. It is sufficient to pick $\xi^\mu=\delta^\mu_i$, a broken translation, to identify the relevant operators, so from now on we consider the equation
\begin{equation}
\braket{\partial_i\phi(w)}=\braket{\phi(w)\, \int_\Sigma d\Sigma_\mu T^\mu_i }~.
    \label{phiQtrans}
\end{equation}
$d\Sigma_\mu T^{\mu}_{i}$ is in the $(\mathbf{1},\mathbf{q})$ under $SO(p)\times SO(q)$, so no defect operator with parallel indices appears in the DOE. While representations of $ SO(q)$ different from the vector appear, they contribute schematically as
$d\Sigma_\mu T^{\mu}_{i} \sim (\alpha\, \delta_i^{j_1}+\beta\, y_i y^{j_1}) y^{j_2}\dots y^{j_s} \hat{O}_{j_1\dots j_s}$. Integrating over the $S^{q-1}$ component of $\Sigma$, one sees that only operators with $s=1$ survive. 

\begin{figure}[htbp]
    \centering
    \begin{tikzpicture}[scale=1.2, transform shape]



    \coordinate (A) at (-0.5,0);
    \coordinate (B) at (4,0);
    \coordinate (C) at (6,1.5);
    \coordinate (D) at (1.5,1.5);

    \fill[LightGrey, opacity=0.4] (A) -- (B) -- (C) -- (D) -- cycle;
    \draw[black,thick] (A) -- (B) -- (C) -- (D) -- cycle;
    \node[above left, xshift=1.8 mm] at (B) {$\partial AdS_{d+1}$};

    \coordinate (P) at (2.4,0.65);

    \draw[teal, thick]
        (P) -- ++(0.6,1.1) node[above right] {\!\!\!\textcolor{teal}{$AdS_{p+1}$}};

          \fill[blue] (P) circle (2pt) node[below, yshift=-0.5 mm, xshift=-0.5 mm] { $\mathcal{D}$}; 

    \draw[dashed]
        (P) -- ++(1,0);

    \draw[->, thick]
        ($(P)+(0.55,0)$)
        arc[start angle=0, end angle=60, radius=0.55];

    \node at ($(P)+(0.75,0.24)$) {$\theta$};

    \draw[->, thick] (O) -- ++(0,1) node[left] {$z$};

    \draw[<-, thick]
        ($(P)+(-0.42,-0.12)$)
        .. controls ($(P)+(-0.62,0.10)$) and ($(P)+(-0.42,0.34)$) ..
        ($(P)+(-0.02,0.26)$);

    \node at ($(P)+(-0.82,0.2)$) {$\varphi_\alpha$};

    \end{tikzpicture}
    \caption{A representation of the $AdS_{p+1}$ slicing coordinate system. 
    }
    \label{fig:ray-angle}
\end{figure}
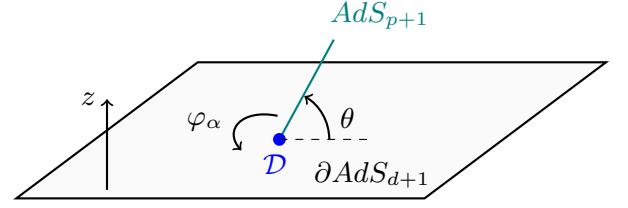

We are led to consider the contribution of the conformal family of a primary $\hat{O}_i$, with scaling dimension $\hat{\Delta}_{\hat{O}}$, to the DOE of the stress tensor. It is convenient to move to the following $AdS_{p+1}$-slicing coordinates for $AdS_{d+1}$
\begin{equation}\label{eq:newcoords}
   \begin{split}
       & \rho=\sqrt{z^2+|y|^2}\,, \quad \theta = \arctan\Big( \frac{z}{|y|}\Big)\,, \quad \varphi_\alpha\,,  \quad x_a\,, \\
   & ds^2=\frac{d \rho^2+dx^2}{\rho^2 \sin^2 \theta}+\frac{d \theta^2}{\sin^2 \theta}+ \cot^2 \theta \, \gamma_{\alpha \beta} d\varphi^\alpha d\varphi^\beta\,.
   \end{split}
\end{equation}
Here, $|y|^2= \delta_{ij}y^i y^j.$ The $AdS_{p+1}$ slices, fibered over the interval $\theta \in [0,\pi/2]$, share their boundaries, at $\rho=0$, where the defect lies. The boundary of $AdS_{d+1}$ is also reached at $\theta=0$, and $\gamma_{\alpha \beta}$ is the metric of the unit $(q-1)$-sphere whose normal vectors are $\hat{y}^i = y^i/|y|$. The correlator $\braket{T_{\mu\nu} \hat{O}_i}$ is determined by the isometries up to seven functions of $\theta$, which appear in the DOE. 
The counting can be performed using a conformal frame \cite{Kravchuk:2016qvl}.
The components of the DOE appearing in $d\Sigma_\mu T^{\mu}_i$ are
\begin{subequations}
\begin{align}
    &T_{\rho \rho} \sim \hat{y}^i\,\big(\mathcal{N}_{2,0} (\theta)+\mathcal{N}_{0,0}^{(1)} (\theta) \big) \bigg(\rho^{\hat{\Delta}_{\hat{O}}-2}\hat{O}_i(x)+\dots \bigg)\,, \\
     &T_{\theta \rho} \sim \hat{y}^i\, \mathcal{N}_{1,0}(\theta) \bigg( \rho^{\hat{\Delta}_{\hat{O}}-1} \hat{O}_i(x)+\dots \bigg)\,, \\
     &T_{\alpha \rho} \sim \partial_\alpha \hat{y}^i\, \mathcal{N}_{1,1}(\theta) \bigg( \rho^{\hat{\Delta}_{\hat{O}}-1} \hat{O}_i(x)+\dots \bigg)\,.
\end{align}
\label{DOEinQ}
\end{subequations}
Here, 
the dots hide the descendants, whose contribution is fixed by the isometries: no function of $\theta$ appears in their coefficient relative to the primary. The notation for the undetermined functions $\mathcal{N}_{a,b}(\theta)$ is related to the spins of the components of $T_{\mu\nu}$ under the little group of a point in the bulk, $SO(p+1)\times SO(q)$. Their dependence on the exchanged operator is not explicit but is understood.
The functions $\mathcal{N}_{a,b}(\theta)$ have power-law asymptotic as $\theta \to 0$, fixed by the BOE of the stress tensor. These limits are reported in the End Matter, section \ref{section:DOEstress}, together with the leading term in the DOE of all the components of the stress tensor. It can be checked that, in the absence of relevant or marginal scalars in the DOE, the unitarity bounds guarantee that the integral \eqref{Qxi} converges. A neutral scalar with $\Delta<d$ must be fine tuned, and depending on the regularization choice, this leads to extra boundary terms in the definition of the charges $Q_\xi$. These contributions precisely subtract the divergence, and one can show that this can be done preserving the topological nature of the charges \cite{Meineri:2023mps}. The net result is simply that one should regulate the integral, subtract the divergence and keep the finite part. In the marginal case, the logarithmic divergence cannot be fine tuned in general \cite{Witten:2001ua,Kaplan:2009kr}: for simplicity, we assume that no marginal operator appears in the BOE of $T_{\mu\nu}$ \footnote{This assumption also excludes primary scalars with $d-\Delta$ equal a positive even integer, which contribute marginal descendants.}.

All in all, by plugging eqs. \eqref{DOEinQ} in \eqref{phiQtrans}, one gets


\begin{equation}
 \braket{\phi(w) \, \int_\Sigma d\Sigma_\mu T^\mu_i }\Big|_{\hat{O}} = \sqrt{\mathcal{C}_{\hat{O}}}\, \rho^{\hat{\Delta}_{\hat{O}}-p-1}  \int d^p x \,  \langle\hat{O}_i(x) \phi(w)\rangle  \,,
\label{phiQxiO}
\end{equation}
where the correlator is restricted to the contribution of one conformal family---but all descendants integrate to zero---and
\begin{multline}
\sqrt{\mathcal{C}_{\hat{O}}} =  \frac{\Omega_{q}}{q}\!\int_0^{\pi/2}\!\!d\theta \,\frac{\cos ^q \theta}{ \sin^{d-1}\theta} \tilde{\mathcal{N}}_{\hat{O}}(\theta)~, \\
\tilde{\mathcal{N}}_{\hat{O}}(\theta)=\bigg(\mathcal{N}_{2,0} (\theta)+\mathcal{N}_{0,0}^{(1)} (\theta) 
- \tan \theta\,\mathcal{N}_{1,0}(\theta)+ \\
(q-1) \mathcal{N}_{1,1}(\theta)/\cos^2(\theta) \bigg)\,.
\label{NO}
\end{multline}
We performed the integral over $S^{q-1}$ which yields the factor $\Omega_q=2\pi^{q/2}/\Gamma(q/2)$. Minor modifications are needed when $q=1$, since an interface may separate two different boundary CFTs: we address this case in section \ref{section:q=1}. Now, the conservation equation $\nabla^\mu T_{\mu\nu}=0$ imposes three differential constraints on the seven functions $\mathcal{N}_{a,b}(\theta)$, reported in (\ref{conseq:p}-\ref{conseq:theta}) in the Supplemental Material. The following linear combination is especially important for us:
\begin{multline}
  (\hat{\Delta}_{\hat{\mathcal{O}}}-p-1)\frac{\cos ^q \theta}{ \sin^{d-1}\theta}\tilde{\Nm}_{\hat{\mathcal{O}}}(\theta) = \\
  \frac{d}{d\theta}\bigg( \frac{\cos ^q \theta}{ \sin^{d-1}\theta} \Big[-\Nm_{1,0}(\theta)-(q-1) \frac{\Nm_{0,1}(\theta)}{\cos^2 \theta}+ \tan \theta \Nm_{0,0}^{(2)}(\theta)\Big]\bigg)\,,
\label{consGood}
\end{multline}
where $\Nm_{0,1},\,\Nm_{0,0}^{(2)}$ are defined in \eqref{eq:completeDOE}. 
For generic $\hat{\Delta}_{\hat{O}}$, therefore, the integrand in \eqref{NO} is a total derivative. Under the assumptions of continuity in AdS and conformality of the boundary theory, the total derivative vanishes at $\theta=0,\,\pi/2$---see sections \ref{section:DOEstress} and \ref{section:q=1} in the End Matter for the precise statement. Hence, $\mathcal{C}_{\hat{O}}=0$ if $\hat{\Delta}_{\hat{O}} \neq p+1$, consistent with the fact that the right hand side of \eqref{phiQtrans} is independent of $\rho$. On the other hand, when $\hat{\Delta}_{\hat{O}} = p+1$, \eqref{consGood} does not constrain $\tilde{\mathcal{N}}_{\hat{O}}$. In fact, we know from \eqref{phiQtrans} that the result cannot vanish. \emph{We conclude that a primary defect operator $\hat{D}_i$, with the quantum numbers of the displacement, exists, and satisfies}
\begin{equation}
    \int_\Sigma d\Sigma_\mu \xi_\nu T^{\mu \nu} = \sqrt{\mathcal{C}_{\hat{D}}} \int_{\mathcal{D}} \xi^i \hat{D}_i~,
    \label{dispWard}
\end{equation}
as an operator equation. We recall that $\mathcal{D}$ is the defect surface at $\rho=0$ and $\mathcal{C}_{\hat{D}}$ is defined by (\ref{DOEinQ},\ref{NO}), with $\hat{O}$ chosen to be the displacement.
The fact that \eqref{dispWard} is valid for any Killing vector can be argued in various ways, for instance noticing that the contribution of the displacement to $\nabla^\mu T_{\mu i}$ provides a representation of a delta function---see \eqref{dispDiffWard}---as we derive in section \ref{section:conservationequations} of the Supplemental Material. Alternatively, one can of course repeat the computation in the general case, or simply take the $\rho \to 0$ limit in \eqref{phiQtrans}, which suppresses the contribution of descendants \footnote{We are ignoring the contribution of other operators, which might contract derivatives of the Killing vector and hence be invisible to the analysis of broken translations. These would appear as descendants or multiply derivatives of a delta function in eq. \eqref{dispDiffWard}
. Since the displacement guarantees that the Ward identities are satisfied, these operators are not required and therefore non-generic.
}.

\section{Examples}
\label{sec:examples}

One can verify the presence of the displacement in many examples. The case of a Wilson loop $W$ in a gauge theory is treated in detail in \cite{Gabai:2025hwf}. Wilson lines can be pushed to the boundary if the gauge field is allowed to fluctuate on $\partial$AdS. In particular, at large $N$, one-point functions of boundary operators normalized by $\braket{W}$ scale like $N^{-1}$, while $\mathcal{C}_{\hat{D}}$ scales like $N^{0}$.

The double-trace interface described in \cite{Melby-Thompson:2017aip} provides another instructive example. On the two sides of the interface, the same free scalar has either of the two boundary conditions allowed by the bulk equations of motion. The resulting defect has a displacement operator. 

In section \ref{sec:examplesSM} of the Supplemental Material, we discuss four other examples. In two of them, the displacement operator exists:
\begin{itemize}
    \item A defect obtained by coupling a free massive boson to a $p$-dimensional Generalized Free Field on the boundary of AdS, following \cite{Bianchi:2024eqm};
    \item The pinning field defect \cite{Allais:2014fqa,Cuomo:2022xgw} in the long-range $O(N)$ model \cite{Bianchi:2024eqm}, at leading order in $d=4-\varepsilon$ expansion. The tilt operator is present as well \cite{Stroppiana2026}.
\end{itemize}

Two other defects for a free massive boson in AdS have no displacement operator and each highlights the optimality of one of the assumptions of the theorem:
\begin{itemize}
    \item A defect that extends in the AdS interior, modeled by a source for the free scalar localized on an AdS$_{p+1}$ slice. In this case, the conservation of the stress tensor is violated in the bulk;
\item A modification of the theory on the boundary of AdS by a non-constant source, following \cite{Herzog:2019bom}. The modification preserves $SO(p+1,1)\times SO(q)$, but extends over the whole $\partial$AdS. In this case, the BOE of $T_{\mu\nu}$ is modified away from the defect, and $Q_\xi$ depends on $\hat{\Sigma}$ non-topologically.
\end{itemize}


\section{Conclusions}
\label{sec:conclusions}

Let us first comment on the implications of our result for the physics of confinement in AdS. On one hand, the existence of the displacement operator puts the study of string excitations in AdS via the corresponding Ward identities on solid ground. On the other hand, our proof shows that the displacement is generic, and in this sense it is not an order parameter for confinement. Any QFT in AdS with a local defect on the boundary supports this excitation, including, for instance, Abelian gauge theories. Only screened defects \cite{Aharony:2023amq}---\emph{i.e.}\ defects invisible to local bulk and boundary operators---have no displacement operator. This implies, for instance, that string breaking in gauge theories with appropriate matter is realized as screening in the boundary CFT. One way to interpret the generic presence of the displacement is that the flux tube always exists at the scale of the curvature. This suggests that better order parameters might be hidden in the spacetime dependence of expectation values or in the scattering amplitudes of the Goldstone bosons, which should exhibit two dimensional dynamics for a confining theory up to scales parametrically larger than the AdS scale. 

It is perhaps useful to comment explicitly on the codimension one case. An interface on $\partial$AdS which is not attached to a brane of course obeys the theorem. Such a defect can never be factorizing: the displacement which obeys \eqref{dispWard} with the unique bulk stress tensor appears in the boundary OPE of operators with a vev on both sides of the interface, whose connected correlator, therefore, does not vanish. Vice versa, boundary conditions are always associated with explicit breaking of the bulk isometries, typically via an end-of-the-world brane \cite{Karch:2000gx,Takayanagi:2011zk}. At the other extreme, an interface which can be tuned to be topological can only exist in a theory with a marginal operator, or extend in the bulk non-trivially \cite{Komatsu:2025cai} \footnote{The double-trace interface described in section \ref{sec:examples} \cite{Melby-Thompson:2017aip} becomes topological only at the BF bound. In this case, the argument in \cite{Komatsu:2025cai} correctly predicts the existence of a marginal operator which does not need to be exactly marginal.}.

 
Turning our attention away from AdS, the proof presented here, with minimal modifications, shows the existence of a displacement operator on conformal defects embedded in other conformal defects in generic CFTs. 
Under the same assumptions as reference \cite{Komatsu:2025cai}, this implies that continuous families of defects in a fixed CFT are generated by marginal defect operators, unless the defect is attached to (topological) surfaces. In the latter case, the interface between two defects in the same family can extend in the bulk as a (non-topological) defect on the topological surface: this happens, for instance, in the case of a monodromy defect \cite{Billo:2013jda,Kravchuk:2025evf}.

Our proof applies to defects in long-range CFTs reached in the IR from a local relevant deformation of Generalized Free Fields \cite{Fisher:1972zz,Paulos:2015jfa, Bianchi:2024eqm}. Indeed, these theories are equivalent to free fields in AdS with boundary couplings. One may ask whether protected operators associated to broken symmetries exist for defects in non-local CFTs without a local higher-dimensional embedding. Intuitively, these operators reflect the ability to couple the defect to the appropriate background fields in a local way, \emph{i.e.} such that their variation inserts the corresponding local operator in the path integral \cite{McAvity:1995zd,Billo:2016cpy}. The role of the higher-dimensional theory with its conserved currents and its local DOE is to give a sharp criterion for this possibility. 

Finally, the fact that the displacement multiplet alone accounts for the breaking of all relevant isometries is the rigorous AdS counterpart to the inverse Higgs constraint in flat space \cite{Volkov:1973vd,Ogievetsky:1974ab,Ivanonv_Ogievetskii,Low_Manohar,Watanabe_Murayama}. It would be interesting to explore whether, in the flat space limit, our construction can provide an equally rigorous proof of the existence of Goldstone modes that propagate in $p+1$ dimensions, without assuming the presence of a $p+1-$dimensional classical surface as in the standard CCWZ treatment \cite{CCWZ,Aharony:2011gb}.

\begin{acknowledgments}

We would like to thank Victor Gorbenko and Petr Kravchuk for useful discussions, Fabiana De Cesare and Lorenzo Di Pietro for comments on the draft, Jiaxin Qiao for coordinating the arxiv submission of our manuscripts, and the organizers and participants of the workshops ``QFT in AdS'' (Trieste 2025 and Lausanne 2026) and ``Interfaces and Symmetry'' (Kyoto 2026) where ideas related to this work originated or were discussed. LB, EdS and MM are supported by the INFN “Iniziativa Specifica” ST\&FI. LB's research is partially supported
by the MUR PRIN contract 2022N9CTAE "Constraining strongly coupled quantum field theories using symmetry". LB and MM are supported by the European Union through the MSCA Staff Exchange program (project HORIZON-MSCA-2023-SE-01-101182937-HeI). MM is supported by the Italian Ministry of University and Research (MUR) under the FIS grant BootBeyond (CUP: D53C24005470001) and the Rita Levi Montalcini program.

\end{acknowledgments}

\newpage

\section*{End matter}
\label{section:endmatter}

\section{The DOE of the stress tensor}
\label{section:DOEstress}

The leading contribution of a defect vector primary to the components of the stress tensor $T_{\mu \nu}$ is
\begin{equation}\label{eq:completeDOE}
    \begin{split}
         &T_{\theta a} \sim  \hat{y}_i\,\mathcal{N}_{1,0}(\theta) \bigg(\frac{\rho^{\hat{\Delta}_{\hat{O}}}}{\hat{\Delta}_{\hat{O}}}\partial_a\hat{O}^i(x)+\ \dots \bigg)\,, \\
     &T_{\theta \rho} \sim \hat{y}_i\, \mathcal{N}_{1,0}(\theta) \bigg( \rho^{\hat{\Delta}_{\hat{O}}-1} \hat{O}^i(x)+\dots \bigg)\,, \\
          &T_{\theta \theta} \sim \hat{y}_i\,\mathcal{N}_{0,0}^{(2)} (\theta) \bigg(\rho^{\hat{\Delta}_{\hat{O}}}\hat{O}^i(x)+\dots \bigg)\,, \\
           &T_{a \rho} \sim \hat{y}_i\, \Big(\tfrac{p+1}{p}\mathcal{N}_{2,0}(\theta) \Big)\bigg(\frac{\rho^{\hat{\Delta}_{\hat{O}}-1}}{\hat{\Delta}_{\hat{O}}}\partial_a\hat{O}^i(x)+\ \dots \bigg)\,, \\
     &T_{ab} \sim \hat{y}_i\, \delta_{ab}\, \Big(\mathcal{N}_{0,0}^{(1)}(\theta) - \tfrac{1}{p} \mathcal{N}_{2,0}(\theta)\Big)\bigg( \rho^{\hat{\Delta}_{\hat{O}}-2} \hat{O}^i(x)+\dots \bigg)\,, \\
          &T_{\rho \rho} \sim \hat{y}_i\,\big(\mathcal{N}_{2,0} (\theta)+\mathcal{N}_{0,0}^{(1)} (\theta) \big) \bigg(\rho^{\hat{\Delta}_{\hat{O}}-2}\hat{O}^i(x)+\dots \bigg)\,, \\
            & T_{\alpha \theta } \sim \partial_\alpha \hat{y}_i \,\mathcal{N}_{0,1} (\theta) \bigg(\rho^{\hat{\Delta}_{\hat{O}}}\hat{O}^i(x)+\dots \bigg)\,, \\
         &T_{\alpha \beta } \sim \gamma_{\alpha \beta} \,\hat{y}_i \,\mathcal{N}_{0,2} (\theta) \bigg(\rho^{\hat{\Delta}_{\hat{O}}}\hat{O}^i(x)+\dots \bigg)\,,\\
         &T_{\alpha a} \sim  \partial_\alpha \hat{y}_i\,\mathcal{N}_{1,1}(\theta) \bigg(\frac{\rho^{\hat{\Delta}_{\hat{O}}}}{\hat{\Delta}_{\hat{O}}}\partial_a\hat{O}^i(x)+\ \dots \bigg)\,, \\
     &T_{\alpha \rho} \sim \partial_\alpha \hat{y}_i\, \mathcal{N}_{1,1}(\theta) \bigg( \rho^{\hat{\Delta}_{\hat{O}}-1} \hat{O}^i(x)+\dots \bigg)\,.
    \end{split}
\end{equation}
These expansions depend on $\theta$ only through the seven functions $\Nm_{a,b}(\theta)$. The ratios of the coefficients of descendants and primary are fixed and independent of $\theta$, as the DOE \eqref{eq:completeDOE} is a standard BOE on each $AdS_{p+1}$ slice.
The combination of stress-tensor components relevant to the Ward identity \eqref{phiQxiO} for orthogonal translations is $T_{i\rho}$ in mixed components, which is explicitly
\begin{equation}
    T_{i\rho} = \hat{y}_i \Big( \cos \theta T_{\rho\rho}-\frac{\sin \theta}{\rho}T_{\theta\rho}\Big)+ \frac{\partial^\alpha \hat{y}_i}{\rho \cos\theta}  T_{\alpha\rho}\,.
\end{equation}
We now derive asymptotic estimates for the functions $\Nm_{a,b}(\theta)$ as $\theta \to 0$, using consistency between the $\theta \to 0$ limit at finite $\rho$ (equivalently, $z \to 0$ at finite $|y|$) and the BOE away from the defect.
The BOE of $T_{\mu\nu}$ away from the defect
can exchange boundary scalars ($\Om$) and rank two symmetric traceless primaries ($\Sm$) of any dimension, or a boundary vector of dimension $\Delta=d+1$. The latter is absent if there is no energy flux across $\partial$AdS, and indeed the existence of conformal charges \eqref{Qxi} requires it. We assume $\Delta_\Om > d$---see comments in the main text for the opposite case. As for $\Sm$, it can be checked that, if it is at the unitarity bound, it does not appear in the BOE. All in all,
\begin{equation}
    \begin{aligned}
      T_{zz} &= \text{O}(z^{d-2+\varepsilon})\,, &   T_{az} &= \text{O}(z^{d-1+\varepsilon})\,,  &
         T_{iz} &= \text{O}(z^{d-1+\varepsilon})\,, \\
         T_{ab} &= \text{O}(z^{d-2+\varepsilon})\,, &
         T_{ij} &= \text{O}(z^{d-2+\varepsilon})\,, &
         T_{ai} &= \text{O}(z^{d-2+\varepsilon})\,.
    \end{aligned}
\end{equation}
for some $\varepsilon>0$. 
In turn, this implies the following asymptotic estimates for the functions $\Nm_{a,b}(\theta)$ as $\theta\to 0$:
\begin{equation}\label{eq:Nthetaasymptotic}
    \begin{aligned}
 \Nm_{1,0}(\theta) &= \text{O}(\theta^{d-1+\varepsilon})\,, 
 &   
 \Nm_{2,0}(\theta) &= \text{O}(\theta^{d-2+\varepsilon})\,, \\
\Nm_{0,1}(\theta) &= \text{O}(\theta^{d-1+\varepsilon})\,, 
   &    
\Nm_{0,2}(\theta) &= \text{O}(\theta^{d-2+\varepsilon})\,,\\  
\Nm_{0,0}^{(1)}(\theta) &= \text{O}(\theta^{d-2+\varepsilon})\,, 
& 
\Nm_{0,0}^{(2)}(\theta) &= \text{O}(\theta^{d-2+\varepsilon})\,,\\
\Nm_{1,1}(\theta) &= \text{O}(\theta^{d-2+\varepsilon})\, . &&
    \end{aligned}
\end{equation}
The estimates \eqref{eq:Nthetaasymptotic} guarantee the convergence of the $\theta$-integral in \eqref{NO} and the vanishing of the boundary term in \eqref{consGood} as $\theta \to 0$.
It is also possible to check that, at $\theta = \pi/2$, there are no convergence problems in \eqref{NO}, and that the boundary term in \eqref{consGood} vanishes, by assuming that $\langle \phi T_{\mu \nu} \rangle$ expressed in Cartesian coordinates is finite at $|y| = 0$ and $z \neq 0$ and continuous everywhere else. This condition is strengthened for $q = 1$, as 
discussed in Section \ref{section:q=1}.


\section{Convergence of the DOE of $Q_\xi$} \label{section:DOEforQ}

In this section, we show that the correlator $\braket{\phi\,Q_\xi}$ can be computed by integrating the DOE of the stress tensor term by term. While the DOE converges absolutely in the region of integration, this is not sufficient in order to exchange the $\rho$-expansion and the $\theta$-integral: a sufficient condition is met if the integral of the series of absolute values converges (Fubini-Tonelli theorem). More directly, we will use the dominated convergence theorem to bound the partial sums. The integral along the parallel directions $x_a$ is harmless and we discuss it last.

The required property can be shown to hold by means of the Cauchy-Schwarz inequality. Since the details of the $SO(d)$ representation of $T_{\mu\nu}$ are irrelevant to the argument and clutter notation, we will replace the stress tensor with a bulk scalar operator $\Phi$. Hence, consider the following correlator and its DOE:
\begin{equation}
    \begin{split}
        & \braket{\phi(w')\,\Phi(x,\hat{y},\rho,\theta)}  =
    \sum_{\hat{O}_n} \mathcal{N}_{\hat{\Delta}_n}(\theta) \rho^{\hat{\Delta}_n} \braket{\phi(w')\,\hat{O}_n(x,\hat{y})}\,, \\ & \hat{O}_n(x,\hat{y}) \equiv 
    \hat{y}^{i_1}\dots \hat{y}^{i_{s_n}}\hat{O}_{n,i_1\dots i_{s_n}}(x)\,, 
    \label{DOEforphiO}
    \end{split}
\end{equation}
where $\rho<1,$ $\hat{O}_n$ are defect scaling operators and $\phi$ is a spectator that lies at $\rho'>1$. With an abuse of notation, the argument of $\Phi$ is written in the coordinate system \eqref{eq:newcoords} while $w'$ denotes Poincaré coordinates as in the main text. As usual, the terms in the $\rho$-expansion are in one-to-one correspondence with coefficients in the expansion of the state $\ket{\Phi(w)}$ on the surface at $\rho=1$ in a complete orthonormal set. The partial sums are, therefore, all bounded by the product of the norms as follows
\begin{equation}
  \begin{split}
      & \left|S_N\right|\equiv \Big| \sum_{\substack{\hat{O}_n \\ n<N}} \mathcal{N}_{\hat{\Delta}_n}(\theta) \rho^{\hat{\Delta}_n} \braket{\phi(w')\,\hat{O}_n(x,\hat{y})} \Big|
     \\
   & \leq \left(\braket{\Phi(x,\hat y,1/\rho,\theta)\Phi(x,\hat{y},\rho,\theta)}\braket{\phi(w')\phi(\mathcal{I}w')}\right)^{1/2}~,
  \end{split}
   \label{PSumBound}
\end{equation}
where $\mathcal{I}w'$ is the result of an inversion centered at the point $w_0=(x^a,y^i=0,z=0)$, \emph{i.e.} $(\mathcal{I}w')^\mu= (w'^\mu-w_0^\mu)/(w'-w_0)^2+w_0^\mu.$ In particular, $(\mathcal{I}w'-w_0)^2=(\rho')^{-2}$ lies inside the unit sphere. Since the right hand side of \eqref{PSumBound} is $N$ independent, the hypotheses of the dominated convergence theorem are satisfied if the two-point function $\braket{\Phi(x,\hat{y},1/\rho,\theta)\Phi(x,\hat{y},\rho,\theta)}$ is integrable in $\theta$ with the appropriate measure. 

One then notices that the small $\theta$ limit of such correlator is 
\begin{equation}
\braket{\Phi(x,\hat{y},1/\rho,\theta)\Phi(x,\hat{y},\rho,\theta)} \sim \theta^{2\Delta}~,
\end{equation}
where $\Delta$ is the scaling dimension of the leading boundary operator in the BOE of $\Phi$. We conclude that integrability of the right hand side of \eqref{PSumBound} is guaranteed if the original correlator \eqref{DOEforphiO} is integrable. In particular, we know this to be the case for the stress tensor (or for an improved version, following \cite{Meineri:2023mps}, when $\Delta<d$). Convergence of the integral over the transverse sphere $S^{q-1}$ is obvious.

Regarding the integral over the directions parallel to the defect, we notice that the dependence of $\braket{\phi(w') \Phi(w)}$ on the AdS$_{p+1}$ coordinates is restricted by isometries to the combination 
\begin{equation}
    \zeta = \frac{\rho \rho'}{(x-x')^2+\rho^2+\rho'{}^2}~.
    \label{zetaCross}
\end{equation}
Therefore, the $\rho$ expansion converges well at large $(x-x')^{-2}$. To prove that it is legal to integrate term by term, we simply bound the tail of the series and show that its integral vanishes in the limit. One can integrate in $\theta,\,\hat{y},\,x$ in any order, so we show the swapping property directly for the $x$ integral of the correlator:
\begin{equation}
    \begin{split}
          & \left| \int d^p x \braket{\phi\, \Phi} - \int d^px\, S_N \right| \leq \\ 
          & \leq K(\theta,\hat{y}) \rho^{\hat{\Delta}_N}
   \left| \int d^p x \left(x^2 +\rho'{}^2\right)^{-(\hat{\Delta}_N+\kappa)/2}\right| \\
  & = K'(\theta,\hat{y})\, \frac{(\rho')^{p-\hat{\Delta}_N-\kappa}}{(\hat{\Delta}_N+\kappa)^{p/2}} \left(1+\text{O}\big(\hat{\Delta}_N^{-1}\big)\right)
   \overset{N\to \infty}{\longrightarrow} 0~.
    \end{split}
\end{equation}
Here, $K,\,K',\,\kappa$ are positive. To obtain the inequality, we used that the DOE converges absolutely to bound the tail by the leading term $\propto \rho^{\Delta_N}$. The function $K$ can be taken independent of $\rho$ and $(x-x')$, by bounding the tail with the sum of absolute values evaluated at $x=x'$ and at a fixed $\rho_0,$  $1>\rho_0>\rho.$ The presence of the positive constant $\kappa$ can be deduced from the cross ratio \eqref{zetaCross} or from a direct analysis of the DOE.

Of course, one should still check that each term integrates to a finite value, and this fails if there are defect operators of dimension $\hat{\Delta} \leq p/2$. Notice that this issue concerns finitely many contributions to the DOE, and is separate from swapping integration and $\rho$-series. When applied to the two-point function of interest $\braket{\phi T_{\rho i}}$, 
a divergence would mean that the charge $Q_\xi$ is not defined. That this does not happen can be seen as follows. Firstly, descendants in the DOE of $T_{\rho i}$ have $\hat{\Delta} > p/2+1$ by unitarity \footnote{The strict inequality follows from the fact that $\Box \hat{O}$ is null if $\hat{O}$ is at the unitarity bound.}, so their leading contribution to the series is $O(x^{-p-\varepsilon})$ and can be integrated. For the primaries, we first cut-off the integral to a compact region in $\mathbb{R}^p$. As explained in the main text, the integral over $\theta$ of the contribution of primaries with dimension $\hat{\Delta} \neq p+1$ vanishes. Therefore, less irrelevant primaries do not contribute to the flux on any surface $\Sigma$ of the kind in figure \ref{fig:tunnel}, cut-off along $\mathbb{R}^p$. Removing the cut-off, we see that the charge is finite \footnote{By changing coordinates so that the defect manifold $\mathcal{D}$ becomes a sphere, one can trace the issue to the pinching of the integration contour on the defect at a point on the sphere. Since the charge is topological, the pinching can be avoided and the result must be finite.}.


\section{Comments on the codimension one case} \label{section:q=1}

The case of an interface, $q=1$, is degenerate, as the sphere $S^{q-1}$ collapses to a set of two distinct points $\{1,-1\}$.
However, the analysis of the previous sections still holds with minor modifications.
First of all, the stress tensor no longer has components in the $SO(q)$-charged indices $\alpha$, and the corresponding functions $\Nm_{a,b}(\theta)$ with $b \neq 0$ must be set to zero.
The integral formula \eqref{NO} is still valid, provided that one removes the prefactor $\Omega_q/q$ and extends the domain of integration to $[0,\pi]$; note that, in the integrand, the $SO(q)$-charged degrees of freedom decouple automatically.
Equation \eqref{consGood} is also valid, as also there $SO(q)$-charged degrees of freedom decouple.
In this case, to discard the boundary term in equation \eqref{consGood} we need $\langle \phi T_{\mu\nu}\rangle$ to be continuous also at $\theta = \frac{\pi}{2}$ rather than just finite, reflecting the fact that a discontinuity on a codimension one surface precisely leads to a delta function contribution in $\nabla^\mu T_{\mu\nu}$.
Finally, if there is reflection symmetry $x \to -x$ (or $\theta \to \pi - \theta$), then the formula \eqref{NO} holds as it stands, provided that $\Omega_q / q \to 2$.
Note that in this case $\Nm_{0,0}^{(2)}(\theta)$ must be odd under reflection and therefore vanish at $\theta = \pi/2$, so that it does not contribute to the boundary term in \eqref{consGood}.

\newpage

\begin{widetext}

\begin{center}
\vspace{1.2cm}
{\Large\bf \mathversion{bold}
{Supplemental material}\\
}
\end{center}

\section{Conservation equations and the differential Ward identity}
\label{section:conservationequations}

The seven functions $\Nm_{a,b}(\theta)$ of the invariant angle $\theta$ that appear in the DOE \eqref{eq:completeDOE}  are constrained by the conservation equation $\nabla^\mu T_{\mu \nu}=0$.
There are three independent constraints, corresponding to $\nu = (a, \rho)$, $\nu = \alpha$, and $\nu = \theta$.
Explicitly, for $\nu =(a, \rho) $ it is
\begin{equation}
    \begin{split}
        &\sin ^2(\theta ) \bigg(\Nm_{1,0}(\theta ) \Big(p \tan (\theta )-\frac{d-1}{\sin (\theta ) \cos (\theta )}\Big)+\Nm_{1,0}'(\theta )+\hat{\Delta}_{\hat{D}}\lsp  \Nm_{0,0}^{(1)}(\theta )+ \Nm_{2,0}(\theta ) (\hat{\Delta}_{\hat{D}} -p-1)\bigg)+ (1-q) \tan ^2(\theta )\Nm_{1,1}(\theta )=0\,.
    \end{split}
    \label{conseq:p}
\end{equation}
For $\nu = \alpha$, it is
\begin{equation}
    \begin{split}
        \tan (\theta ) \Nm_{0,1}(\theta ) \left(p-\frac{d-1}{\sin ^2(\theta )}\right)+\Nm_{0,1}'(\theta )+\frac{\Nm_{0,2}(\theta )}{\cos ^2(\theta )}+\Nm_{1,1}(\theta ) (\hat{\Delta}_{\hat{D}} -p)=0\,.
    \end{split}
    \label{conseq:q}
\end{equation}
For $\nu= \theta$, it is
\begin{equation}
    \begin{split}
        &\sin ^2(\theta ) \bigg(\Nm_{0,0}^{(2)}(\theta ) \Big((1-q) \tan (\theta )-(d-2) \cot (\theta )\Big)+\Nm_{0,0}^{(2)}{}'(\theta )+(p+1) \cot (\theta ) \Nm_{0,0}^{(1)}(\theta )+ \\
        & \hspace{3 cm}+\frac{(1-q) \Nm_{0,1}(\theta )}{\cos ^2(\theta )}+(\hat{\Delta}_{\hat{D}}-p)  \Nm_{1,0}(\theta )\bigg)+ \frac{(q-1) \sin (\theta )\Nm_{0,2} (\theta )}{\cos ^3(\theta )}=0\,.
    \end{split}
    \label{conseq:theta}
\end{equation}
To derive the distributional identity \eqref{dispDiffWard}, we
use the DOE \eqref{eq:completeDOE}
with $\hat{\Delta}_{\hat{O}}=p+1+\varepsilon$,
and we compute the divergence
\begin{equation}
    \begin{split}
    \label{eq:distributional1}
         &  \sqrt{g}\, \nabla^\mu T_{\mu i} = \partial_\rho (\rho^\varepsilon)  H_i(\theta,\varphi,x)+\text{O}(\rho^\varepsilon)\,,
  \\
 & H_i(\theta,\varphi,x) = \sqrt{\gamma}\Big(
\mathcal{G}_{\parallel}(\theta)\,\hat y^i \hat y^j
+
\mathcal{G}_{\perp}(\theta)\,
(\delta^{ij}-\hat y^i \hat y^j)
\Big)
\hat{O}_j(x)  \,, \\
    &    \mathcal{G}_{\parallel}(\theta)
=
\frac{\cos^q\theta}{\sin^{d-1}\theta}
\left(
\mathcal{N}_{2,0}
+
\mathcal{N}^{(1)}_{0,0}
\right)
-
\frac{\cos^{q-1}\theta}{\sin^{d-2}\theta}
\mathcal{N}_{1,0}\,, \\
& \mathcal{G}_{\perp}(\theta)
=
\frac{\cos^{q-2}\theta}{(\sin^{d-1}\theta)}
\mathcal{N}_{1,1}\,.
    \end{split}
\end{equation}
Given a smooth test function $f(\rho,\theta,\varphi)$ supported for $\rho \leq R$ and regular at the boundary, we can integrate it against $\nabla^\mu T_{\mu i}$
\begin{equation}\label{eq:divergencetest}
  (\sqrt{g}\,\nabla^\mu T_{\mu i }\,|\, f) \equiv \lim_{\varepsilon\to 0}  \int_{\rho >0} \!\! d\rho d \theta d \Omega_{q-1} f(\rho,\theta,\varphi) \sqrt{g}\, \nabla^\mu T_{\mu i} =  \int d\theta d\Omega_{q-1} \big(\lim_{\rho \to 0 }f(\rho,\theta,\varphi)\big) H_i(\theta,\varphi,x)+ \text{O}(R)\,,
\end{equation}
where in the second equality we used \eqref{eq:distributional1} and the fact that $\partial_\rho(\rho^\varepsilon)$ is a representation of a delta function on the half-line \cite{gelfand1964vol1}.
The distribution $\nabla^\mu T_{\mu i}$ is supported at $\rho=0$ since it is vanishing for every $\rho >0$. Therefore, we can take $R$ to be arbitrarily small, and the remainder term in \eqref{eq:divergencetest} must be zero.
Finally, $\lim_{\rho \to 0} f(\rho,\theta,\phi)=f(x,y)\big|_{z=y=0}=f_0$ is just a number since $f$ is regular at the boundary, and the remaining angular integral is exactly the same as in \eqref{NO}. We obtain
\begin{equation}
(\sqrt{g}\,\nabla^\mu T_{\mu i }\,|\, f) = f_0\sqrt{C_{\hat{D}}} \hat{D}_i(x)\,,
\end{equation}
where we replaced $\hat{O}_i$ with $\hat{D}_i$ since their dimensions match at $\varepsilon=0.$
Since $f_0 = (\delta(z)\delta^q(y) | f)$, we find the distributional identity \eqref{dispDiffWard}.

\section{AdS conserved current and the tilt operator} \label{tilt}

Consider a QFT in AdS with a global symmetry associated to the conserved current $J^\mu(w)$. As we have done for the displacement, we will show the existence of a protected tilt operator on a conformal defect which breaks the global symmetry at the AdS boundary. For simplicity of notation, we consider a $U(1)$ subgroup of the symmetry group. When the CFT on $\partial AdS$ contains a conserved current, it sources a gauge field in the AdS interior. We assume this is not the case and that the symmetry is non-locally realized on the boundary, \emph{i.e.}\
it is generated by a charge operator that extends inside AdS. Technically, this assumption translates in the absence of a scalar operator in the BOE of the current $J^{\mu}$ away from the defect. Indeed, 
a scalar operator in the BOE of an AdS conserved current is allowed only when the scalar has dimension $\Delta_\mathcal{O}=d$ and this implies a non-vanishing flux of the current through the AdS boundary. Assuming this does not happen, the $U(1)$ symmetry is only broken at the location of the defect.

As in the main text, we will use the Poincaré coordinate system described below equation \eqref{dispDiffWard} and the slicing coordinates \eqref{eq:newcoords}. The components of the current $J^{\mu}$ in the two coordinate systems are related by
\begin{equation} \label{eq:currentchangecoords}
    \begin{split}
         &J_\theta = \rho \Big( \cos \theta J_z - \sin \theta \,\hat{y}^i J_i \Big)\,, \\
       & J_\rho =  \sin \theta J_z + \cos \theta \,\hat{y}^i J_i \,,\\
       & J_{\alpha}=\rho \cos \theta \partial_\alpha \hat y^i J_i\,,
    \end{split}
\end{equation}
 and $J_a$ is not affected by the change of coordinates. We consider the scalar contribution to the DOE of the current in the coordinates \eqref{eq:newcoords}, where we can use the $AdS_{p+1}$ isometries to fix
\begin{equation}\label{eq:OPEdefectirho}
    \begin{split}
        &J_a \sim \mathcal{N}_1(\theta) \bigg(\frac{\rho^{\hat{\Delta}_{\hat{O}}}}{\hat{\Delta}_{\hat{O}}}\partial_a\hat{O}(x)+\ \dots \bigg)\,, \\
          &J_\rho \sim \mathcal{N}_1(\theta) \bigg( \rho^{\hat{\Delta}_{\hat{O}}-1} \hat{O}(x)+\dots \bigg)\,,\\
           &J_\theta \sim \mathcal{N}_2 (\theta) \bigg(\rho^{\hat{\Delta}_{\hat{O}}}\hat{O}(x)+\dots \bigg)\,.
    \end{split}
\end{equation}
$\mathcal{N}_1(\theta)$ and $\mathcal{N}_2(\theta)$ are normalization factors that depend on the dynamics of the theory. 
Note that the component $J_\alpha$ has no scalar contribution in its BOE as the preserved isometries impose that $J_i \propto \hat{y}_i$ is purely radial and $\hat{y}^i \partial_\alpha \hat y_i=0$.
We can also combine \eqref{eq:currentchangecoords} and \eqref{eq:OPEdefectirho} to write $J_i$ and $J_z$ explicitly
\begin{equation}\label{eq:defectOPEza}
    \begin{split}
        &J_z \sim  \mathcal{N}_z(\theta)\rho^{\hat{\Delta}_{\hat{O}}-1}\hat{O}(x)+\dots \,,\\
        &J_i \sim \hat{y}_i \,
\mathcal{N}_\perp(\theta)\rho^{\hat{\Delta}_{\hat{O}}-1}\hat{O}(x)+\dots  \,,
    \end{split}
\end{equation}
where
\begin{equation}\label{eq:NzNpasymptotics}
    \begin{split}
          &\mathcal{N}_z(\theta) \equiv \sin \theta \,\mathcal{N}_1(\theta)+ \cos \theta \, \mathcal{N}_2(\theta)\,, \quad \mathcal{N}_\perp(\theta) \equiv \cos \theta \,\mathcal{N}_1(\theta)- \sin \theta \, \mathcal{N}_2(\theta)\,.
    \end{split}
\end{equation}
A differential constraint on the functions $\mathcal{N}_1(\theta)$ and  $\mathcal{N}_2(\theta)$ is provided by the conservation of the AdS current which imposes
\begin{equation}\label{eq:diffeqN}
   (p-\hat{\Delta}_{\hat{O}})\frac{\cos^{q-1}\theta}{\sin^{d-1}\theta}\mathcal{N}_1(\theta)=\frac{d}{d\theta} \left(\frac{\cos^{q-1}\theta}{\sin^{d-1}\theta}\mathcal{N}_2(\theta)\right)\,.
\end{equation}
As we have done in the main text, we will consider the charge
\begin{equation}
Q=\int_{\Sigma} d\Sigma_{\mu} J^{\mu}~, 
\label{Q}
\end{equation}
and its action on a charged scalar operator $\phi(w)$ lying on $\partial AdS$ away from the defect
\begin{equation}
    \braket{\delta\phi(w)}=\braket{\phi(w)\, \int_\Sigma d\Sigma_\mu J^\mu }\,.
    \label{phiQJ}
\end{equation}
Using the DOE of the current $J^{\mu}$, we can compute the contributions of each exchanged defect operator to \eqref{phiQJ}. While the current exchanges defect operators with $SO(p)$ spin one and multiple representations charged under $SO(q)$ \cite{Lauria:2018klo}, only defect scalars survive the integral \eqref{Q}. For the latter, we obtain
\begin{equation}
\left. \braket{\phi(w) \, \int_\Sigma d\Sigma_\mu j^\mu }\right|_{\hat{O}} = \sqrt{\mathcal{C}_{\hat{O}}}\, \rho^{\hat{\Delta}_{\hat{O}}-p}  \int d^p x \,  \langle\hat{O}(x) \phi(w)\rangle  \,,
\label{phiQO}
\end{equation}
with
\begin{equation}\label{eq:CO}
\sqrt{\mathcal{C}_{\hat{O}}}=\Omega_{q-1}\!\int_0^{\pi/2}\!\!d\theta \frac{\cos^{q-1} \theta}{\sin^{d-1}\theta} \mathcal{N}_1(\theta)\,.
\end{equation}
Notice that, for $\hat{\Delta}_{\hat{O}}\neq p$, the conservation equation \eqref{eq:diffeqN} imposes that the integrand is a total derivative of a function that vanishes for $\theta=0$ (as $\mathcal{N}_2(\theta)\sim \theta^{\Delta_{\mathcal{J}}}$ for $\theta \to 0$ where $\mathcal{J}$ is a vector operator in the BOE of the current $J$ and $\Delta_{\mathcal{J}}>d-1$ for the unitarity bound of the boundary CFT) and $\theta=\pi/2$ (see section \ref{section:q=1} for the case $q=1$). The only 
way to obtain a non-vanishing contribution is to consider a scalar operator of dimension $\hat{\Delta}_{\hat O}=p$. In that case, the convergence of the integral \eqref{eq:CO} is guaranteed by the asymptotic behavior $\mathcal{N}_1(\theta)\sim \theta^{\Delta_{\mathcal{J}-1}}$ for $\theta\to 0$. Therefore, symmetry breaking implies the existence of a defect scalar operator of protected dimension $p$, which is the \emph{tilt} operator.

\section{Details on the examples}
\label{sec:examplesSM}

\subsection{Defects with displacement operator}

\subsubsection{A free theory example}

A simple example of a defect in a non-local CFT is obtained by coupling a generalized free field (GFF) in $d$ dimensions to another GFF that lives on a $p$-dimensional linear subspace.
Consider the action
\begin{equation}\label{eq:defectaction1}
     S= \int d^du \, \tfrac{1}{2} \phi S^d_{2\Delta_\phi}\phi+  \int d^p x \, \Big( \tfrac{1}{2}  \hat \psi S^p_{2\Delta_{\hat \psi}} \hat \psi+  h\,\phi^n  \hat \psi^m \Big)\,,
\end{equation}
where $u=(x^a,y^i)$ are parallel and orthogonal directions to the $p$-dimensional defect, $m$ and $n$ are positive integers, and
\begin{equation}
    S^d_{\alpha}\mathcal{O}(u) \equiv \kappa^d_\alpha \int d^dv  \frac{\mathcal{O}(u)}{|u-v|^\alpha}\,, \quad \quad  \kappa^d_\alpha= \frac{2^{d-\alpha}\lsp\Gamma\big( d-\frac{\alpha}{2}\big)}{\pi^{\frac{d}{2}} \lsp \Gamma \big( \frac{d-\alpha}{2}\big)}\,.
\end{equation}
If in \eqref{eq:defectaction1} we  take $ n \Delta_\phi+m  \Delta_{\hat \psi}=p-\varepsilon$ with $\varepsilon \ll 1$ for some specific choices of integers $n$ and $m$, this action has an IR fixed point at $h_* \propto \varepsilon$ or $h_* \propto \varepsilon^{1/2}$ that corresponds to a defect CFT \cite{Bianchi:2024eqm}.
For instance, to get a real fixed point one can take $d=3$, $p=1$ and $(n,m)=(1,2)$, or $d=3$, $p=2$ and $(n,m)=(2,2)$.

The dCFTs at these fixed points arise naturally as boundary theories of local QFTs in AdS. Indeed,
GFF theory is the boundary limit of a free massive scalar field $\Phi$ in AdS
\begin{equation}
    S_{\text AdS_{d+1}}= \int_{z>0} \frac{ dz \lsp d^du}{z^{d+1}} \left ( \frac{1}{2} \partial_\mu \Phi \partial^\mu \Phi + \frac{1}{2}M^2 \Phi^2\right)\,,
    \label{SfreeAdS}
\end{equation}
where the mass $M$ of the AdS field and the conformal dimension $\Delta_\phi$ of the GFF are related through $M^2 = \Delta_\phi(\Delta_\phi - d)$. If we replace $\phi(u)$ with $\Phi(u,z)$
while keeping the field $\hat{\psi}$ and the interaction term on the $p$-dimensional defect on $\partial AdS$, 
we are only modifying the free theory \eqref{SfreeAdS} on the defect. If we decide to also lift $\hat{\psi}$ to a free scalar in $AdS_{p+1}$, we see that the coupling of the two theories is local, at the defect, hence clearly the DOE of operators in AdS$_{d+1}$ is local. 
In this way, all the hypotheses of the theorem are satisfied, and it follows that a displacement operator must exist.
In fact, it is possible to directly show to all orders in $\varepsilon$ that the following operator is protected: 
\begin{equation}
    \tilde D_i(x)= \hat \phi^{n-1}\partial_i  \hat \phi \,\hat \psi^m(x)\,,
\end{equation}
where $\hat{ \phi}(x)=\phi( x,y=0)$. We have
\begin{equation}
    \tilde D_i(x) = \sqrt{\Nm} \hat{D}_i(x)~,
\end{equation}
where $\hat{D}_i$ is unit-normalized and $\Nm$ is a constant computable in perturbation theory.
Classically, the dimension of this operator is $n \Delta_\phi + m \Delta_{ \hat \psi} = p + 1 - \varepsilon$, but it receives an anomalous dimension that exactly cancels the $\varepsilon$ contribution.
This is due to a factorization property that holds to all orders in perturbation theory:
\begin{equation}\label{eq:factorisationtwopt}
    \langle \tilde{D}_i(x) \tilde{D}_j(x') \rangle = \langle \partial_i \hat{\phi}(x) \partial_j \hat{ \phi}(x ')  \rangle  \,\langle \hat{\phi}^{n-1} \hat \psi ^m(x) \hat{\phi}^{n-1} \hat \psi ^m(x')  \rangle\,.
\end{equation}
The simple reason for this factorization is that all the fields in the interaction term in \eqref{eq:defectaction1} do not carry any $SO(q)$ index.
Any diagram in which $\partial_i \hat{\phi}(x)$ is connected to the interaction through a free propagator $G_0(u)$ vanishes, since $\partial_i G_0(x,y)\big|_{y = 0} = 0$.
Therefore, in any non-vanishing diagram, $\partial_i \hat{\phi}(x)$ and $\partial_j \hat{\phi}(x')$ must be connected by a free propagator.

For the same reason, the operator $ \partial_i \hat{\phi}(x) $ turns out to be protected with dimension $\hat{\Delta}_{\partial_i \hat{\phi}} = \Delta_\phi + 1$, as its two-point function only has the tree-level diagram.
Finally, the operator $\hat{\phi}^{n-1} \hat{\psi}^m(x)$ also has protected dimension $\hat{\Delta}_{\hat{\phi}^{n-1} \hat{\psi}^m} = p - \Delta_\phi$, as is argued in \cite{Bianchi:2024eqm} both from the equation of motion and from the defect-channel decomposition of the bulk two-point function $\langle \phi(u) \phi(v) \rangle$.
Then \eqref{eq:factorisationtwopt} implies $\hat{\Delta}_{\hat D} = \hat{\Delta}_{\partial_i \hat{\phi}} + \hat{\Delta}_{\hat{\phi}^{n-1} \hat{\psi}^m} = p + 1$. \\

In this example, one can also derive the modified Ward identity for the stress tensor explicitly.
The boundary field $\phi(u)$ is the leading operator in the BOE of the AdS field $\Phi(u,z),$
\begin{equation}\label{eq:PhiOPE}
    \Phi(u,z) \sim c_\phi \ z^{\Delta_\phi} \phi(u)+\ldots\,. 
\end{equation}
If we regulate the theory by cutting off AdS at $z=\eta>0$, the equation of motion for $\Phi(u,z)$ is
\begin{equation}\label{eq:EOMPhi}
    \sqrt{g} \big(\Box_{\text AdS}-M^2\big)   \Phi(x,y,z) = \frac{\delta S^\eta_{\text{defect}}}{\delta  \Phi(x,y,z)} 
\end{equation}
where $S^\eta_{\text{defect}}$ is the second term in \eqref{eq:defectaction1}, evaluated on the cutoff surface.
The stress tensor is
\begin{equation}\label{eq:AdSfreestresstensor}
    T_{\mu \nu}= \partial_\mu \Phi \partial_\nu \Phi -\frac{1}{2}g_{\mu \nu}((\partial \Phi)^2+M^2 \Phi^2)\,,
    \end{equation}
    and by using \eqref{eq:EOMPhi} we find
    \begin{equation}
       \sqrt{g} \lsp \nabla^\mu T_{\mu \nu}= \partial_\nu \Phi \frac{\delta S^\eta_{\text{defect}}}{\delta \Phi(x,y,z)}~.
    \end{equation}
Since the theory is free, using \eqref{eq:EOMPhi} and \eqref{eq:PhiOPE} we get
\begin{equation}
    \lim_{\eta \to 0} \Big(\nabla_i \Phi \frac{\delta S^\eta_{\text{defect}}}{\delta \Phi(x,y,z)} \Big) =\frac{n h}{2} \hat \phi^{n-1}\partial_i \hat\phi\, \hat{\psi}^m(x)\,
    \delta(z) \delta^q(y)\,,
\end{equation}
so that, at the fixed point,
\begin{equation}
    \sqrt{g} \lsp \nabla^\mu T_{\mu i}= \frac{n h_*}{2} \tilde D_i(x)\delta(z) \delta^q(y)\,,
    \
\end{equation}
and therefore $\mathcal{C}_{\hat D} = \Nm (nh_*/2)^2$. 
As a sanity check, the same value can also be obtained from a CFT computation via
\begin{equation}
    \sqrt{\mathcal{C}_{\hat D}} = \frac{ \int d^px' \langle \phi^2(x,y) \hat D_i(x') \rangle_\Dm}{\partial_i\langle \phi^2(x,y) \rangle_\Dm}\,.
\end{equation}
The one-point function in the denominator has already been computed in \cite{Bianchi:2024eqm}.
For the bulk-to-defect two-point function in the numerator, there is a factorization property analogous to \eqref{eq:factorisationtwopt} that holds at all orders in perturbation theory and reads
\begin{equation}
    \langle \phi^2(x,y) \tilde D_i(x') \rangle = \langle \phi(x,y) \partial_i \hat{\phi}(x') \rangle \langle\phi(x,y)\hat \phi^{n-1} \hat{\psi}^{m}(x') \rangle\,,
\end{equation}
and both correlators can be evaluated exactly (up to the determination of $h_*$) using the same techniques as those used for $\langle \phi^2(x,y) \rangle$ in \cite{Bianchi:2024eqm}.
The result is once again $\mathcal{C}_{\hat D} = \Nm(n h_*/2)^2$. \\

\subsubsection{The localized magnetic field in the long-range Ising model}\label{sec:lmfdisplacement}

Starting from GFF, we can add an interaction term
\begin{equation}\label{eq:longrangeaction}
     S_0= \int d^du \, \Big(\tfrac{1}{2} \phi S^d_{2\Delta_\phi}\phi +  \tfrac{\lambda}{4!}  \phi^4\Big)\,.
\end{equation}
If one takes $\Delta_\phi= \frac{d-\delta}{4} $, where $\delta$ is small, the interaction is weakly relevant, and it triggers a renormalization group flow to a conformal fixed point at $\lambda_* \propto \delta$ \cite{Fisher:1972zz,Paulos:2015jfa}.
One can add a line defect to this picture by considering a localized magnetic field
\begin{equation}\label{eq:longrangedefectint}
S = S_0 + h_0 \int dx \, \phi(x,y=0)\,,
\end{equation}
where $x$ is the one dimensional coordinate along the defect and $y^i$ are the $d-1$ dimensional coordinates orthogonal to it.
If we take $d=4-\varepsilon$ and $\Delta_\phi= 1- \frac{\kappa+1}{4} \varepsilon$ \footnote{Here $\Delta_\phi$ is a parameter of the Lagrangian, and should not be confused with the quantum conformal dimension of the operator $\phi$; this is only the classical dimension, and there is a non-zero anomalous dimension (see \cite{Bianchi:2024eqm} for more details).}
for $\varepsilon$ small and $0<\kappa \leq 1$, then both the bulk and defect interaction terms are weakly marginal, and the flow can be studied using perturbation theory \cite{Bianchi:2024eqm}.
The IR fixed point couplings are
\begin{equation}
\lambda_* = \frac{(4 \pi)^2}{3}\kappa\lsp \varepsilon + \text{O}(\varepsilon^2)\,, \quad h_*^2 = \frac{9(\kappa+1)}{2 \kappa} +\text{O}(\varepsilon)\,.
\end{equation}
The value $\kappa = 1$ corresponds to the localized magnetic field defect fixed point in the local Ising model, where a displacement operator exists by the usual Ward identity argument in local defect CFTs.
This theory can be uplifted to an AdS theory by promoting $\phi(x,y)$ to an AdS free field $\Phi(x,y,z)$ of mass $M^2=\Delta_\phi(\Delta_\phi-d)$ as before. The interaction term in \eqref{eq:longrangeaction} is an interaction supported at the boundary that involves the first contribution in the BOE of $\Phi$ as $z \to 0$. Similarly, the defect interaction in \eqref{eq:longrangedefectint} is supported  at the defect location on the boundary ($z=0, y=0$) and does not modify the BOE of any AdS field at $y \neq 0$.
Therefore, the argument given in \ref{sec:generalargument} can be applied to conclude the existence of a displacement operator $\hat{D}_i(x)$. Up to an overall constant $\tilde{D}_i(x) = \sqrt{\Nm}\hat{D}_i(x)$, it must be $\tilde{D}_i(x)= \partial_i \phi(x)$ since this is the only object with dimension $p+1+\text{O}(\varepsilon)$ in the correct representation of $SO(q)$.
We can check at first non-trivial order in perturbation theory that this operator is indeed protected.
Up to order $\varepsilon$, we only need to consider two diagrams for the correlator $\langle \partial_i \phi(x)  \partial_j \phi(x') \rangle$ 
\begin{equation}
 	\begin{tikzpicture}[scale=0.4, baseline]
	 \draw[thick,blue] (-3,0)--(3,0);
	  \draw[thick, dashed, black]    (-2,0) to[out=90,in=180] (0,2);
	   \draw[thick, dashed, black]    (2,0) to[out=90,in=0] (0,2);
	  	   \node[below] at (-2,0) {$\partial_i \phi $};
	  	   \node[below] at (2,0) {$\partial_j \phi$};
	\end{tikzpicture}
	 \hspace{1 cm}
	  	\begin{tikzpicture}[scale=0.4, baseline]
	 \draw[thick,blue] (-3,0)--(3,0);
	 \draw[thick, dashed, black]    (-2,0) to[out=90,in=180] (0,2);
	   \draw[thick, dashed, black]    (2,0) to[out=90,in=0] (0,2);
	   \draw[thick, dashed, black]    (0.66,0)--(0,2);
	    \draw[thick, dashed, black]    (-0.66,0)--(0,2);
  \draw[black, fill=black]   (-0.66,0) circle (4pt);
\draw[black, fill=black] ($(0,2)+(-4pt,-4pt)$) rectangle ($(0,2)+(4pt,4pt)$);
	  	  \draw[black, fill=black]   (0.66,0) circle (4pt);
	  	   \node[below] at (-2,0) {$\partial_i \phi $};
	  	   \node[below] at (2,0) {$\partial_j \phi$};
	\end{tikzpicture}
 \end{equation}
where the black box and circle represent the bulk and defect interaction, respectively, and the dashed line is the free propagator obtained from the kinetic term in \eqref{eq:longrangeaction}.
The second diagram is divergent, and this divergence must be reabsorbed by the wavefunction renormalization factor $(\partial_i \phi(x))_R = Z^{-1}_{\partial_i \phi} (\partial_i \phi(x))_0$, where $R$ ad $0$ denote the renormalized and the bare operator, respectively.
In the MS scheme in dimensional regularisation one finds
\begin{equation}
Z_{\partial_i \phi}= 1 - \frac{1}{48(3\kappa+1)\pi^2} \frac{\lambda h^2}{\varepsilon}+ \text{O}(\lambda^2)\,.
\end{equation}
The anomalous dimension at the fixed point is
\begin{equation}
\gamma_{\partial_i \phi}= \Big[\beta_\lambda \frac{\partial \log Z_{\partial_i \phi}}{\partial \lambda} +\beta_h \frac{\partial \log Z_{\partial_i \phi}}{\partial h} \Big]_{\lambda_*,h_*}= \frac{\kappa+1}{4}\varepsilon+\text{O}(\varepsilon^2)\,,
\end{equation}
where for this computation it is sufficient to use the classical part of the beta functions $\beta_\lambda = - \kappa \varepsilon \lambda + \text{O}(\lambda^2)$ and $\beta_h = - \frac{\kappa +1}{4} \varepsilon h + \text{O}(\lambda)$.
The anomalous dimension is exactly the one needed to get a protected operator at this order, indeed
\begin{equation}
\Delta_{\partial_i \phi}= 2-\frac{\kappa+1}{4}\varepsilon+\gamma_{\partial_i \phi}= 2+\text{O}(\varepsilon^2)\,,
\end{equation}
independently of $\kappa$, as expected. The same results hold also for the long-range $O(N)$ model and the same localized magnetic field defect, with minor modifications.
In this case, the defect also breaks $O(N)$ symmetry to $O(N-1)$, and consequently there is an exactly marginal tilt operator corresponding to each of the $N-1$ broken generators \cite{Stroppiana2026}, as argued in Section \ref{tilt}.

\subsection{A defect without a displacement operator}

There is another way to construct a non-trivial defect in GFF in flat space: one can again start from the GFF action
\begin{equation}
     S= \int d^du \, \tfrac{1}{2} \phi S^d_{2\Delta_\phi}\phi\,,
\end{equation}
and then add a $p$-dimensional conformal defect by turning on a nonzero one-point function as follows
\begin{equation}\label{eq:GFFonepoint}
    \langle \phi(x,y)\rangle = \frac{ a_\phi}{|y|^{\Delta_\phi}}\,.
\end{equation}
Since the theory is quadratic, all correlators are defined through Wick contractions of \eqref{eq:GFFonepoint} and the free bulk propagator.
This can also be achieved by adding to the action an external source coupled to $\phi$:
\begin{equation}\label{eq:GFFdefectaction}
    S_{\text{source}}= h \int d^px d^q y  \, \frac{\phi(x,y)}{|y|^{d-\Delta_\phi}}\,,
\end{equation}
where $h \propto a_\phi$ \footnote{The proportionality constant diverges for $\Delta_\phi = p$. In that case, the one-point function \eqref{eq:GFFonepoint} is produced by the defect $\int d^{p}x\lsp\phi(x)$, which is the (free limit of) the pinning field defect and has a displacement operator.}.
Although this term is not localized on the defect, it still preserves the defect symmetry group.
In this example, the defect spectrum is trivial: there are no anomalous dimensions, so all defect operators have dimension $\Delta_\Om = m \Delta_\phi + n$, with $m,n$ integers. 
In particular, for generic $\Delta_\phi$, there is no displacement operator, as there are no operators with dimension $p+1$.
In this simple model, the absence of a displacement operator is due to the fact that the defect interaction in \eqref{eq:GFFdefectaction}, although preserving the defect symmetry group, is not truly localized on the defect but rather spread throughout space. \\

A natural question is what the AdS counterpart of this example is, and why the argument given in the main text cannot be applied to infer the existence of a protected displacement operator.
It turns out that there are two inequivalent AdS realizations of this boundary non-local defect CFT. They fail to satisfy the assumptions of the theorem for two different reasons. In one case, the stress tensor is not conserved everywhere in the bulk, while in the other the BOE is modified away from the defect, leading to a non-topological charge $Q_\xi$.
For both examples, the starting point is the AdS dual of GFF \eqref{SfreeAdS}. \\

\paragraph{First realization.}
To reproduce the correlators of the boundary CFT with the source \eqref{eq:GFFdefectaction}, we can add an interaction term localized on a brane $\text{AdS}_{p+1} \subset \text{AdS}_{d+1}$ defined by fixing the $q$ coordinates $y^i = 0$. This brane breaks the AdS isometries down to a group isomorphic to the symmetries of the defect theory. Specifically, we add to the action the term
\begin{equation}
    S_{\text{brane}} = g \int_{z>0} \frac{dz \, d^p x}{z^{p+1}} \, \Phi(x,y=0,z)\,,
\end{equation}
with $g \propto h$. Since the theory is quadratic, the only non-trivial observable in the boundary CFT is the one-point function $\langle \phi(x,y) \rangle$, where $\phi(x,y)= \lim_{z \to 0} (z^{-\Delta_\phi}\Phi(x,y,z))$.
This is given by a single Witten diagram that can be easily computed \cite{Gimenez-Grau:2023fcy}
\begin{align}\label{eq:wittendiag1}
& \langle \phi(x,y) \rangle \ =  \ \begin{tikzpicture}[scale=0.5, baseline=-0.1cm]
 \draw[fill= white, draw=black, line width=0.25mm] (1.5,0) circle (42pt);
    \draw[thick,  blue] (2,1.4) to (2.,-1.4);
     \draw[thick, dashed, black] (0,0) to (2,0);
       \draw[black, fill=black] (2,0) circle (3pt);
       \node[left] at (0,0) {$(x,y)$};
	\end{tikzpicture} 
	\hspace{0.2cm} =   \ g \int_{z'>0} \frac{dz' \, d^p x'}{(z')^{p+1}} \, K_{\Delta_\phi}(x-x',y,z')\,,
\end{align}
where $K_\Delta(x,y,z)=\big(\frac{z}{|x|^2+|y|^2+z^2} \big)^\Delta$ is the bulk-to-boundary propagator.
The result is
\begin{equation}
    \langle \phi(x,y) \rangle = \frac{\pi^\frac{p}{2}\,\Gamma(\frac{\Delta_\phi}{2})\,\Gamma(\frac{\Delta_\phi-p}{2})\lsp g}{2\lsp \Gamma(\Delta_\phi) }\cdot\frac{1}{|y|^{\Delta_\phi}}\,,
    \label{onepointphiWitten}
\end{equation}
which can be matched to \eqref{eq:GFFonepoint}.
It is also interesting to compute the one-point function of the fundamental field in the interior of AdS, $\langle \Phi(x,y,z)\rangle$. This is given by a diagram similar to \eqref{eq:wittendiag1}, but with the external point inserted in the bulk instead of on the boundary, or equivalently, by replacing the bulk-to-boundary propagator in the integral with a bulk-to-bulk propagator.
Since the integral obtained in this way is quite involved, we instead use a trick to get to the result.
By applying the equation of motion to the one-point function we have
\begin{equation}\label{eq:EOMonepoint}
    \big(\Box_{\text AdS}-M^2\big) \langle \Phi(x,y,z)\rangle = g\lsp z^{q}\lsp\delta^{q}(y)\,.
\end{equation}
Moreover, the preserved AdS symmetries impose that 
\begin{equation}
    \langle \Phi(x,y,z)\rangle= f \bigg( \frac{|y|^2}{z^2} \bigg)\,.
\end{equation}
Using this in \eqref{eq:EOMonepoint}, we obtain a second order ordinary differential equation for $f(\chi)$, where $\chi = |y|^2 / z^2$
\begin{equation}
    4 \chi(1+\chi)f''(\chi)+2(q+(d+2)\chi)f'(\chi)-\Delta_\phi(\Delta_\phi-d) f(\chi)=0\,.
\end{equation}
This equation has two independent solutions; the correct linear combination is fixed by imposing the appropriate boundary conditions.
The most general solution is \footnote{
For $q = 2$, these two solutions become degenerate, and the second independent solution is a more complicated function. However, the same reasoning still applies.
}
\begin{equation}
    \begin{split}
      f(\chi)=  \alpha \, \chi^\frac{2-q}{2} \,{}_2 F_1 \bigg( \frac{p+2-\Delta_\phi}{2}, \frac{2-q+\Delta_\phi}{2}, \frac{4-q}{2}; -\chi\bigg) + \beta\, {}_2F_1 \bigg( \frac{d-\Delta_\phi}{2}, \frac{\Delta_\phi}{2}, \frac{q}{2}; -\chi \bigg)\,.
    \end{split}
    \label{generalsol}
\end{equation}
The boundary condition at $\chi=0$ comes from the delta-function behavior in \eqref{eq:EOMonepoint}. To reproduce it, one must have 
\footnote{
This is the same behavior one would have in flat space. In fact, the computation differs from that in flat space only by curvature corrections, which are suppressed at short distances.
}
\begin{equation}
    f(\chi) \overset{\chi \to 0}{\sim } \chi^{\frac{2-q}{2}}\,.
\end{equation}
This holds provided that $\alpha \neq 0$.
On the other hand, as $z \to 0$ ($\chi \to \infty$), we must have
\begin{equation}
    f(\chi) \overset{\chi \to \infty}{\sim } \chi^{-\frac{\Delta_\phi}{2}}\,,
\end{equation}
and in particular there cannot be terms with asymptotic behavior $\chi^{\frac{\Delta_\phi-d}{2}}$, as can be derived from the Witten diagram representation and the fact that the GFF spectrum does not contain operators with dimension $d-\Delta_\phi$. 
This implies the condition
\begin{equation}
    \frac{ \Gamma\big(\frac{4-q}{2}\big)}{\Gamma\big(\frac{\Delta_\phi+2-d}{2}\big)\Gamma\big(\frac{\Delta_\phi+2-q}{2}\big)}\,  \alpha \,+ \, \frac{\Gamma \big( \frac{q}{2}\big)}{\Gamma \big( \frac{\Delta_\phi}{2} \big)\Gamma \big( \frac{\Delta_\phi-p}{2} \big)}\, \beta \, = \,0\,.
    \label{alphabeta}
\end{equation}
Finally, the overall proportionality constant can be easily determined by matching to \eqref{onepointphiWitten} in the limit $z \to 0$.
In this theory, the presence of the brane explicitly breaks the conservation of the stress-energy tensor in the interior of AdS; thus, the argument given in the main text cannot be applied, and specifically the boundary term at $\theta=\pi/2$ in \eqref{consGood} does not vanish.\\

\paragraph{Second realization.}
Consider the same free massive scalar field on AdS.
This time, we want to implement the presence of the defect through a boundary condition.
Specifically, we set
\begin{equation}\label{eq:boundarysource}
   \big(z^{\Delta_\phi-d} \Phi(x,y,z)   \big)\big|_{z=0}= \frac{h}{|y|^{d-\Delta_\phi}}
   \,.
\end{equation}
The boundary condition acts as an external source according to the AdS/CFT dictionary \cite{Gubser:1998bc,Witten:1998qj} and is exactly the same as in \eqref{eq:GFFdefectaction}. 
In this example as well, it is interesting to consider the one-point function of $\Phi(x,y,z)$.
Since the theory is quadratic, this is the only relevant observable.
From the bulk-to-boundary propagator one gets
\begin{equation}\label{eq:sourceintegral}
    \langle \Phi(x,y,z)\rangle=h \, z^{\Delta_\phi} \int \frac{d^p x' d^q y'}{|y'|^{d-\Delta_\phi}(z^2+|x-x'|^2+|y-y'|^2)^{\Delta_\phi}}\,.
\end{equation}
The integral converges for $ \Delta_\phi>p$, and it can be evaluated in closed form and yields
\begin{equation}\label{eq:AdSonepoint2}
   \langle \Phi(x,y,z)\rangle\propto h \, {}_2F_1 \bigg( \frac{d-\Delta_\phi}{2}, \frac{\Delta_\phi}{2}, \frac{q}{2}; -\frac{|y|^2}{z^2}\bigg)\,.
\end{equation}
Due to the boundary condition \eqref{eq:boundarysource}, the BOE of $\Phi(x,y,z)$ is 
\begin{equation}
    \Phi(x,y,z) \sim    \Big( z^{\Delta_\phi}\phi(x,y)+\ldots\Big)+ \bigg(\frac{h }{|y|^{d-\Delta_\phi}}z^{d-\Delta_\phi}+\ldots\bigg)\,.
    \label{BOEPhisecond}
\end{equation}
Combining this with the $z\to 0$ limit of \eqref{eq:AdSonepoint2}, one sees that the one-point function of $\phi$ \eqref{eq:GFFonepoint} is correctly reproduced. However, the bulk one-point function \eqref{eq:AdSonepoint2} differs from the first realization, eqs. (\ref{generalsol},\ref{alphabeta}), since in this case $\alpha=0$. In particular,
here the equation of motion is satisfied without any contact term in the AdS interior
\begin{equation}\label{eq:EOMonepoint2}
    \big(\Box_{\text{AdS}} - M^2\big) \langle \Phi(x,y,z) \rangle= 0 \,,
\end{equation}
and the one-point function is nonsingular, taking a finite nonzero value at $y = 0$. 
Moreover, since the theory is quadratic, it follows that the stress-energy tensor is conserved everywhere in \text{AdS}.
However, the argument given in the main text still fails. 
In fact, the external source \eqref{eq:boundarysource} modifies the BOE of the operators of the theory away from the defect (\emph{i.e.}\ for $|y| \neq 0$).
This is most easily seen in the BOE \eqref{BOEPhisecond}, where the second term is absent without the source.  
Using \eqref{eq:AdSfreestresstensor} and the fact that the theory is quadratic, it is easy to see that such terms are also present in the BOE of the stress tensor. 
In particular, $T_{zi}$ receives a contribution of order $z^{d-1}$:
\begin{equation}
    T_{zi} \sim z^{d-1}\, h(d-\Delta_\phi)\,|y|^{\Delta_\phi-d-2} \left(|y|^2 \partial_i \phi-\Delta_\phi\, y_i\, \phi\right) +\dots
    \label{tziflux}
\end{equation}
Here, the dots hide other singular terms which are either c-numbers---and so disconnected in correlation functions---or total derivatives along $y^i$, which can be removed by improving the stress tensor \cite{Meineri:2023mps}. For generic $\Delta_\phi$, the $\text{O}(z^{d-1})$ term in \eqref{tziflux} is not a total derivative, and yields a flux of the translation current through $\partial$AdS. Hence, the corresponding charge is not conserved and the operator $Q_\xi$ in \eqref{Qxi} is not topological. Correspondingly, the boundary term at $\theta=0$ in \eqref{consGood} does not vanish.

\end{widetext}

\bibliographystyle{apsrev4-1}
\bibliography{references}

\end{document}